\documentclass[a4paper,12pt]{article}
\usepackage{epsfig}
\usepackage{amsmath}
\usepackage{amsfonts}
\usepackage{graphics}
\usepackage{graphicx}

\def\beq{\begin{equation}}
\def\eeq{\end{equation}}
\def\bea{\begin{eqnarray}}
\def\eea{\end{eqnarray}}
\def\barr{\begin{eqnarray}}
\def\earr{\end{eqnarray}}

\def\u1{\widehat{U(1)}}
\def\su2{\widehat{SU(2)}_1}

\def\uq2{U_q(sl(2))}
\def\Uq2{\widehat{U_q(sl(2)}}

\def\Z{{\bf Z}}
\def\nl{\nonumber \\}

\def\a{\alpha}

\def\de{\partial}

\begin{document}

\begin{titlepage}

\begin{center}
\hfill  \quad  \\
\vskip 0.5 cm
{\Large \bf Universal power-law exponents in 
differential tunneling conductance for
planar insulators near Mott criticality 
at low temperatures }

\vspace{0.5cm}

Federico~L.~ BOTTESI ,\ \ Guillermo~R.~ZEMBA\footnote{
Fellow of CONICET, Argentina.}\\

{\em Facultad de Ingenier\'ia y Ciencias Agrarias,  Pontificia Universidad Cat\'olica Argentina,}\\
{\em  Av. Alicia Moreau de Justo 1500,(C1107AAZ) Buenos Aires, Argentina}\\
\medskip
{\em and}\\
\medskip
{\em Departamento de F\'{\i}sica, Laboratorio Tandar,}\\
{\em  Comisi\'on Nacional de Energ\'{\i}a At\'omica,} \\
{\em Av.Libertador 8250,(C1429BNP) Buenos Aires, Argentina}
\end{center}
\vspace{.3cm}
\begin{abstract}
\noindent
We consider the low-temperature differential tunneling conductance 
$G$ for interfaces between a planar insulating material in the Mott-class and a metal. 
For values of the the applied potential difference $V$ that are not 
very small, there is a experimentally observed universal regime in which
$G \sim V^m$, where $m$ is a universal exponent. We consider 
the theoretical prediction of the values of $m$ by using the method of 
Effective Field Theory ($EFT$), which is appropriate for discussing universal phenomena. 
We describe the Mott material by the $EFT$
pertaining the long-distance behavior of a spinless Hubbard-like model with nearest 
neighbors interactions previously considered. At the Mott transition, the $EFT$ is
known to be given by a double
Abelian Chern-Simons theory. The simplest realization 
of this theory at the tunneling interface yields a Conformal Field Theory with 
central charges $(c,\bar c) =(1,1)$  
and Jain filling fraction $\nu = 2/3$ describing a pair  
of independent counter-propagating chiral bosons (one charged and one neutral). Tunneling 
from the material into the metal is, therefore, described 
by this $EFT$ at the Mott critical point. The resulting tunneling conductance behaves 
as $G \sim V^{(1/\nu -1)}$, yielding the prediction $m=1/2$, which compares well 
(within a $10 \%$ deviation) with the results for this exponent in two experimental studies considered here. 
\end{abstract}

\vskip 0.5 cm
\end{titlepage}
\pagenumbering{arabic}

Tunneling spectroscopy \cite{tunneling} is a powerful experimental tool to probe response functions of materials,
such as the differential conductance $G$, which is defined as $G = dI/dV$ , where $I$ is the
resulting tunneling current flowing through the junction for the applied potential difference $V$ across it. 
For values of $V$ that are smaller than a characteristic scale $V_0$, $G$ 
encodes a non-universal behavior, but for $V > V_0$ there is a universal regime characterized by
a relation in the form of a 'non-linear Ohm's Law', $ I \sim V^{m+1}$, where $m$ is a universal exponent. 
This relation leads to $G \sim V^m$, characteristic
of this type of experiments.

In this letter, we shall focus on the universal exponent $m$ that characterizes the quantum tunneling  
between 
a planar insulator ({\it e.g.}, doped semiconductors or semiconductor oxides such as $SrTiO_3$) 
near the Mott-like critical point and a conductor, which is the 
situation relevant to two series of experimental results that we will address 
here \cite{Lee-Massey} \cite{Marshall}. 
The universal regime of (moderately) large $V$ can be successfully described by 
Effective Field Theories ($EFT$s) \cite{Polchinski}. 
One such an effective theory pertaining the class of materials considered in the 
experimental 
studies \cite{Lee-Massey} \cite{Marshall}
was proposed in \cite{Botte-Z-Mott1}.
It was obtained as the long-range limit of a discrete planar lattice $AF$ model, near
the critical Mott point. The field content of this theory is given in terms of 
two Chern-Simons Abelian fields. These fields give rise to a Conformal Field Theory ($CFT$) \cite{cft}
at the boundary of the sample. 
For the experimental setups considered here, there is no external magnetic field that would break
the chiral symmetry of the $CFT$, as in the case of the Quantum Hall Effect 
($QHE$). Nevertheless, the results of \cite{Botte-Z-Mott1} indicate a 
factorized chiral structure on this $CFT$. Even beyond the critical point,
the idea of a factorized structure could be pursued. 
In the following, we shall assume that the temperature is low enough so that
its energy scale could be disregarded.

For the sake of completeness, we shall briefly review here the key aspects of this theory 
\cite{Botte-Z-Mott1}. 
We first consider the $EFT$ pertaining a planar Mott material in the bulk and in the tunneling 
boundary. 
We start by considering the following Hamiltonian model on a 2D spatial lattice:
\beq
H_{2D}\ =\ -\frac{t}{2}\ \sum_{x,\mu} [\ \psi^\dagger(x+ a e_\mu) e^{i a_\mu} \psi(x)\ 
+\ {\rm h.c.}\ ]\  +\ U\ \sum_{x,\mu} \rho(x)\rho(x+a e_\mu)\ , \label{Model-Ferm-2d}
\eeq
where $\psi(x)$ is the fermion field, $x$ labels the lattice sites and $e_\mu$ are 
the unit lattice vectors pointing to the nearest 
neighbors of a given site, $a$ is  the lattice spacing ,
$t$ is the hopping parameter, $U$ is the (constant) Coulomb potential, $\rho(x)$ is the  
charge density (normal-ordered with respect 
to the half-filling ground state),  $\rho(x)= [:\psi^{\dagger}(x)\psi(x): -1/2]$ and 
$a_\mu $
is an Abelian statistical gauge field defined on the links of the lattice. 
This model is well suited for describing electrons in systems with `narrow pocket Fermi surfaces' (see, {\it e.g.}, \cite{Botte-Z-Mott1}), relevant for the experimental
setups considered here.
In the continuum (thermodynamic) limit the above model can be mapped onto a two-dimensional anisotropic Heisenberg ($XXZ$-spin model) by means of a 
two-dimensional Jordan-Wigner transformation. The low-energy, long-range $EFT$ of this model yields 
an Abelian gauge theory with two bosonic statistical Chern-Simons fields.

For the sake of clarity of the exposition, let us first consider an Abelian Chern-Simons 
gauge theory, 
based on a single statistical field 
$A_\mu (x)$ defined on $(2+1)$-dimensional space-time ($x$ denotes now  
a $(2+1)$-dimensional coordinate $x^\mu $, with $\mu=0,1,2$  ).
Its action is given by:
\beq
S_{CS}\ =\ \frac{k}{4\pi} \int d^3x \epsilon^{\mu,\nu,\lambda} A_\mu \de_\nu A_\lambda  \quad ,
\label{singleCS}
\eeq
where $k$ is the self-coupling constant.
This  a topological gauge field theory, meaning that the is no bulk dynamics since its
hamiltonian vanishes identically ($H = 0$). 
Subtleties regarding the signature of space-time are overlooked in this simple discussion
(see \cite{Witten}). 
The natural observables of this 
theory are Wilson Loops, or flux tubes:
\beq
W_n[ \Gamma\ ] =\ P\exp{(in\oint_\Gamma A_\mu dx^{\mu})}  \quad , n \in \mathbb{Z} \qquad ,
\label{wilson}
\eeq 
where $P$ denotes path-ordering and the exponents may be thought of as statistical fluxes 
subtended by the Wilson loop $\Gamma$.
Notice that the switching of the sign of the coupling constant $k$ in the action could 
be absorbed by a sign flip of the gauge field, leaving the action unmodified but changing the sign 
of the statistical fluxes \cite{Witten}. 

The $EFT$ for the model (\ref{Model-Ferm-2d}) slightly away from the Mott critical point is given by the 
double lattice Chern-Simons action:
\bea
S_{DCSL}=\frac{k}{4\pi} \int d^3x\ A^{(1)}_\mu K_{\mu \nu} A^{(1)}_{\nu} 
-\frac{k}{4\pi} \int d^3x\  A^{(2)}_\mu K_{\mu \nu}A^{(2)}_{\nu}\ , \label{Double-CS}
\eea 
with coupling constants $k >0$ and $-k$, and where $A^{(1)}$ and $A^{(2)}$ are two Abelian statistical gauge 
fields. 
Here we consider the coordinates $x$ on a cubic lattice of spacing $a$, with
forward difference operators $d_\mu f(x)=[f(x+a\epsilon_\mu)-f(x)]/a$,
forward shift operators $S_\mu f(x)=f(x+a\epsilon_\mu)$, 
and the kernel $K_{\mu \nu}=S_{\mu}\epsilon_{\mu \a \nu}d_\a$ (no summation is implied over
equal indices $\mu$ and $\nu$) \cite{Carlo-Topics}.

Notice that the fields $A^{(1)}$ and $A^{(2)}$ may be thought of as having opposite orientation 
in the Wilson loop circulation, or, equivalently, as having statistical fluxes of opposite sign (consider $n=1$ 
in (\ref{wilson})).
This theory has Quantum Group Symmetry $U_q\left( \hat{sl}(2) \right ) \otimes U_q\left( \hat{sl}(2)\right )$ 
with deformation 
parameter $q=e^{i\pi/k}$. At the critical Mott point, 
$q=-1$ and $k = 1$ \cite{Witten} \cite{grensing}.

In the rest of our analysis, we shall consider that we are close enough to the Mott
critical point so that the $EFT$ may be further simplified with action:
\bea
S_{EFT}\ =\ \frac{1}{4\pi} \int \ d^3x \epsilon^{\mu \nu \lambda} A^{(1)}_\mu \de_\nu A^{(1)}_\lambda 
-\frac{1}{4\pi} \int d^3x\  \epsilon^{\mu \nu \lambda} A^{(2)}_\mu \de_\nu A^{(2)}_\lambda \ ,
\label{EFTaction}
\eea
where the the continuum limit of (\ref{Double-CS}) has been taken. 
This is a theory of two Abelian Chern-Simons statistical gauge fields with 
two types of vortices, given by the opposite spatial 
Wilson loop orientations (defined, {\it e.g.}, with respect to the spatial plane, but 
consistent definitions in Euclidean 3D space can be achieved nonetheless \cite{zemba} ). 
As any topological theory, it may give rise to
dynamical phenomena only at the boundaries of the spatial region. This region 
could be a line or segment. For the sake of simplicity, we consider it to be a circle of
radius $R$ (any boundary will provide a proper length scale). We shall assume that the 
difference in statistical magnetic flux originated
in both fields gives rise to a pair of chiral bosonic fields with different 
propagating senses along this line.
That is, we consider that the difference in Wilson loop vorticity translates into 
two chiralities at the boundary (further discussion on the boundary projection
of Wilson loops may be found in \cite{zemba}).
In this sense, the dynamics becomes identical to that
of the edge states in a sample displaying the Quantum Hall Effect ($QHE$).

In the following, we consider the action (\ref{EFTaction}) to be defined on a space-time
manifold $\cal{M}$ with the topology of a cylinder made from a spatial disk, parametrized
by an angle $\theta$ and a radius $R$, and time $t$. In this geometry, the Chern-Simons 
action induces a boundary field theory with dynamics given by the choice of boundary conditions
\cite{wen} , which may be identified with a $CFT$ \cite{Witten}. The Gauss Law constraint derived 
from (\ref{EFTaction}) imposes flat field strength tensors
$F^{(i)}_{\mu \nu} = 0$, where $F^{(i)}_{\mu \nu} = \partial_\mu A^{(i)}_{\nu}- \partial_\nu A^{(i)}_{\mu}$, 
$(i=1,2)$. The 'pure gauge' fields that solve this constraint are expressed in terms
of bosonic Abelian ones $\varphi^{(i)} $ such that 
\beq
A^{(i)}_\mu\ =\ - \frac{1}{2\pi} \partial_\mu \varphi^{(i)}\ ,
\label{bosonization}
\eeq
where $i=1,2$.
The bosonic fields can be taken as propagating boundary fields, provided adequate 
boundary conditions are chosen (see, {\it e.g}, \cite{wen}). 
We note that global parity is preserved in the action (\ref{EFTaction}), and therefore
we take the bosonic fields as counter-propagating modes at the spatial boundary of $\cal{M}$.
We choose, without loss of generality, $\varphi^{(1)} $ ($\varphi^{(2)} $) to be 
right (left) handed, respectively, {\it i.e}, $\varphi^{(i)}(R\theta \mp v_{i} t)$,
where $v_i$ are positive dimensionless Fermi velocity parameters, introduced by the 
boundary conditions. It is well-known that the free bosonic chiral (anti-chiral) theory 
of $\varphi^{(1)} $ ($\varphi^{(2)} $) defines a $c=1$ (${\bar c}=1$ $CFT$. The latter 
theory is usually defined on the complex plane obtained by mapping of the cylinder onto it
by $z=exp(\tau +i \theta)$, where $\tau = it$ is the Euclidean time. 

In order to make the $CFT$ explicit, we consider the Abelian (chiral and anti-chiral) currents:
\beq
J^{(i)}= - \frac{1}{2\pi} \frac{\partial \varphi^{(i)}}{\partial\theta}
=\frac{1}{2\pi}\  \sum_{n=-\infty}^\infty \ 
\a^{(i)}_n {\rm e }^{in\left( \theta \mp v_i t/R \right)} \ ,\qquad
i=1,2\ ,
\label{currents}
\eeq
whose Fourier modes $\alpha_n^{(i)}$ satisfy the
two-component Abelian current algebra,
\beq
\left[\ \a^{(i)}_n \ ,\ \a^{(j)}_m\ \right] =\delta^{ij}\ n \ 
\delta_{n+m,0} \ .
\label{curralg}
\eeq
The corresponding generators of the Virasoro algebra are obtained by
the familiar Sugawara construction 
$L^{(i)} = : \left( J^{(i)} \right)^2 :/2$, and read:
\beq
L_0^{(i)} = \frac{1}{2} \a^{(i)}_0 + \sum_{n=0}^\infty\ 
\a_{-n}^{(i)} \a_n^{(i)} \ ,\qquad\qquad   L_m^{(i)} = \frac{1}{2} \sum_{n=-\infty}^\infty\ 
\a_{m-n}^{(i)} \a_n^{(i)} \    \qquad i=1,2\ .
\label{ln}\eeq
They satisfy
\beq
\left[\ L_n^{(i)} \ ,\ L_m^{(j)} \ \right] = \delta^{ij}
\left\{ (n-m) L_{n+m}^{(i)} + \frac{c}{12}
n(n^2-1) \delta_{n+m,0} \right\}\ , \qquad c=1\ .
\label{viralg}
\eeq
The generators of conformal transformations are thus given by
$L_n=L^{(1)}_n+L^{(2)}_n $. The highest weight representations of the Abelian
algebra (\ref{curralg}) are labeled by the
eigenvalues of $J^{(i)}_0$ and $L^{(i)}_0$;
the eigenvalue of $\left(J^{(1)}_0+J^{(2)}_0 \right)$ is
proportional to the quasi-particle charge $Q$ and
$2\left( L^{(1)}_0+L^{(2)}_0 \right)$ is their
fractional statistics $\theta/\pi$.
The Hilbert space of the two-component Abelian $CFT$ is made of a consistent set of these representations, 
which is complete with respect to the fusion rules, which are the selection rules for the composition of
quasi-particle excitations in the theory
(the so-called bootstrap self-consistent conditions) \cite{cft}.
In the Abelian theory, the fusion rules simply require the addition of the two-component charges of the 
quasi-particles; as a consequence,
the allowed charge values correspond to the points of a two-dimensional lattice.
Each lattice specifies a theory: the adjacency matrix of the lattice 
contains some parameters which are partially determined by the
physical conditions. These determine the quantization of 
charges and conformal dimensions. 
Special lattices allow for an extended symmetry 
\cite{abe}: the extension from 
$\u1\times\u1$ to $\u1_{\rm diagonal}\times\su2$
is relevant for the present $CFT$, as we shall consider in a moment.
The bosonic field corresponding to the $\u1_{\rm diagonal} $
is assumed to be compactified on a circle, which introduces the compactification radius as 
a parameter that defines the unit of charge. 
The Hamiltonian of the Abelian theory which assigns
a linear spectrum to the edge excitations can be written
in terms of the currents (\ref{currents})
as follows:
\beq
H_R= \frac{\pi}{R^2}\int_0^{2\pi R }  dx : \left( 
v_1 J^{(1)} J^{(1)} - v_2 J^{(2)} J^{(2)} \right):  =
\frac{1}{R} \left[ v_1 L_0^{(1)} - v_2 L_0^{(2)} -\frac{1}{12} \right] \ .
\label{hcft}
\eeq
We emphasize here, however, that the spectrum that is relevant for discussing the Hilbert 
space of the $CFT$ is that of the charge and Virasoro operators rather than that of the
edge Hamiltonian. 

The coupling of the Abelian $CFT$ to the physical
electric field should be addressed
when considering tunneling experiments. 
In terms of the individual currents (\ref{currents}), the electric 
current should be taken as the symmetric 
combination of both, given that the labels $i=1,2$ are arbitrary. 
We therefore introduce the current and Virasoro generators in the basis
that factorizes the theory into charged and neutral sectors \cite{cz}:
\beq
\begin{array}{llcl}
J = J^{(1)} + J^{(2)} \ , & \frac{J_0}{\sqrt{2s-1}} &
\longrightarrow & Q = \frac{2\ell}{2s-1}\ ; \\
J^3 = \frac{1}{2}\left( J^{(1)} - J^{(2)} \right) \ , & J^3_0 & 
\longrightarrow & n \ ;\\
L = L^{(1)} + L^{(2)} = L^Q +L^S \ , & L_0 &
\longrightarrow & \frac{\ell^2}{2s-1} + n^2 \ ;\\
L^Q = \frac{1}{4} \ :\ \left( J\right)^2 \ :\ , \quad 
L^S = \ :\ \left( J^3 \right)^2 \ : \ .& &
\end{array}
\label{coeq}\eeq
We have introduced the charge scale parameter $1/\sqrt{2s-1}$ which is given by the compactification 
radius of the charged bosonic field. 
The corresponding spectrum of (\ref{coeq}) may be determined from the bosonic Fock space structure and 
was found to be 
quantized according to \cite{abe}\cite{cz}: $s=2,4,\dots$, and that it splits in two cases: 
$(I)$: $l \in \mathbb{Z} $, $n \in \mathbb{Z} $, and 
$(II)$: $l \in \mathbb{Z} +1/2$, $n \in \mathbb{Z} +1/2$.
The spectrum yields the values of charge $Q$ and fractional statistics 
$2h = \theta/\pi$ of the quasi-particle excitations ($h$ is the
conformal dimension, {\it i.e.}, the eigenvalue of $L_0$ and $\theta$ the quantum statistical phase).
The discretization of the parameters in (\ref{coeq}) is the consequence of 
imposing upon the spectrum the condition of the existence of
excitations with the quantum numbers of physical electrons, {\it i.e.}, $Q=1,2,\dots$ and $2h = 1,3,\dots$.
{\it A posteriori}, this spectrum is identified with that of the Jain states \cite{Jain}, for opposite moving
chiral bosons \cite{abe} and filling fractions $\nu = 2/(2s-1)$, with  $s=2,4,\dots$. 
Each value in the spectrum (\ref{coeq}) is the highest weight of a pair of representations of the Abelian 
current algebra, which describe 
an infinite tower of edge excitations, generated by the 
bosonic Fock-space operators $\a^{(i)}_n$, $n<0$, $i=1,2$
\cite{cft}, 
corresponding to Fermionic particle-hole transitions, 
or to their anyonic generalizations.

The neutral sector of the spectrum 
displays an extended symmetry $\su2$, which is apparent in this basis.
We observe that there are two highest weights with dimension
one, {\it i.e.} $n=\pm 1$, which correspond to the additional currents:
\beq
J^\pm = \ : \exp \left( \pm i \sqrt{2}\varphi \right):\ ,
\qquad \varphi = \frac{1}{2}\left( \varphi^{(1)} -\varphi^{(2)} \right)\ .
\label{jpm}\eeq
The two fields $J^\pm$, together with $J^3$ in (\ref{coeq}), form the
$\su2$ current algebra of level $k=1$; their Fourier modes satisfy:
\barr
\left[\ J^a_n\ ,\ J^b_m \ \right] &=& i\epsilon^{abc} J^c_{n+m} +
\frac{k}{2} \delta^{ab} \delta_{n+m,0}\ , \quad k=1 ,\ a,b,c =1,2,3,\nl
\left[\ L^S_n\ ,\ J^a_m \ \right] &=& -m \ J^a_{n+m} \ ;
\label{naca}
\earr
note that the operators $J^a_n$ commute with the generators of the charged sector
$\left( L^Q_n, J_m \right)$.
As is well known, there are two highest-weight representations of
the $\su2$ algebra, which are labeled by the isospin $\sigma=\alpha /2$, with $\alpha =0,1$.

The Jain spectrum of the theory defined by (\ref{coeq}) and (\ref{hcft}) can now be
written in this basis that makes apparent the decomposition into $\u1$ and $\su2$ sectors \cite{cz} , 
so that both cases $(I)$ and $(II)$ may be displayed together:
\barr
\nu=\frac{2}{2s - 1}\ ,\qquad 
\ Q & = & \frac{ 2l - \alpha}{2s - 1}\ ,\nl
L_0 &=& \frac{\left(2l -\alpha\right)^2}{4(2s - 1)} - 
\frac{\alpha(2-\alpha)}{4} +\ r\ , \quad r \in{\Z}\ .
\label{usspec}\earr
Here $l \in \mathbb{Z} $.
We verify that, for $\alpha =1$ and $ l = s$ we obtain
$Q=1$ and $2h = s-1 + 2r$, which are the correct quantum numbers for electrons. 

We now turn our attention to the tunneling between the Mott
material and a metal. 
The characteristics of the tunneling current $I$ versus the applied voltage $V$ 
have been studied in the context of Luttinger models, which are Abelian $CFT$s, in 
\cite{Kane-Fisher} \cite{Chamon-Fradkin}. For the sake of completeness,
we provide here an independent derivation of the simplest of their results based on a scaling
argument \cite{Polchinski} (we thank Andrea Cappelli for suggesting this to us). 

The tunneling of a $CFT$ may be described in standard fashion (see \cite{Glattli}) 
at low temperatures: in the experiments it is assumed that the thermal energy is much smaller 
than any other relevant energy scale.
We consider the real coupling of a localized tunneling interaction to the
effective $CFT$ action describing the Mott material
as in \cite{Chamon-Fradkin} and \cite{Glattli}:
\beq
S_T = g \int \Psi^{\dagger}\Psi \delta (x) dxdt ,
\label{tunnelings}
\eeq
where $\Psi (t,x)$ is the free fermion field describing the 
electrons. 
Without loss of generality, we assume the tunneling to be localized on the $1D$ boundary
at $x=0$.

Only physical electrons tunnel between the material described by the $CFT$ and the metal,
so that $\Psi$ has electric charge $Q=1$ and conformal dimension 
$h=p/2$, where $p$ is an odd integer. We first consider the case 
of the Laughlin states, with filling
fraction $\nu=1/p$, in which all the quasi-particles in 
the spectrum are electrically charged. 
Therefore, the scaling dimension of the coupling constant $g$ is $(1-p)$. 
To first perturbative order in $g$, this yields
$I \sim g V^p  $    
because both $V$ and $I$ have scaling dimension $1$ ($I$ is the time derivative 
of the charge, or, equivalently, electrical resistance is dimensionless). 
Therefore, we conclude that  
$dI/dV \sim V^{(p-1)}$ .
The result for the conductance, is, therefore:
\beq
G=dI/dV \sim V^{(1/\nu-1)}\ ,
\eeq 
which predicts $m=(1/\nu-1)$ and that 
was obtained in \cite{Kane-Fisher} \cite{Chamon-Fradkin} using alternative methods. 
Following \cite{Glattli}, we generalize this result to other $CFT$s describing interacting electrons such that 
$\nu$ may take more general values. Those are theories in
which the quasi-particle excitations may have one electrically charged and neutral modes. Nevertheless, 
there is always an excitation with the quantum numbers of the electron among those, which is the only one that 
may tunnel into the metal. 
In these more general scenarios, the tunneling interaction is always
irrelevant in the sense of the Renormalization Group 
\cite{Polchinski} since (minus)
the scaling dimension of the coupling constant $g$ satisfies
$(1/\nu -1)>0$, which holds for any values of $\nu < 1$, {\it i.e.}, all
the ones considered here. It is, however, useful to note that the 
interaction is increasingly irrelevant as $\nu$ takes smaller values.

We now consider the $CFT$ describing the insulator and its 
relation to the tunneling experiments.
As discussed before, the $(c, \bar{c}) = (1,1)$ $CFT$ 
has symmetry $\widehat{U(1)}\times \widehat{SU(2)}_{1}$,
and $\nu =2/(2p - 1)$, $p=2,4,\dots$. The lowest possible
values are $\nu = 2/3, 2/7, \dots$. As noted before,
all of these values yield irrelevant
interactions, but $\nu = 2/3$ is the least  
irrelevant among them, becoming therefore the favored 
value. A similar discussion applies to the Fermi
interaction for the Beta-decay \cite{Polchinski}.
For this filling fraction, the tunneling exponent yields
$m =(1/\nu-1)= 1/2$, and this is therefore the prediction
of our $EFT$.
This value compares well with the relevant experimental data 
of \cite{Lee-Massey} \cite{Marshall}. In the latter,
we consider the experiments done for $Sm$ only, which are the only ones that fit in our 
assumptions for the $EFT$, {\it i.e.}, $AF$ order, insulator. 
In the former, it is reported that $m \approx 0.43-0.47$,
whereas in the second it was found that $m \approx 0.44$. 
The deviation from the theoretical prediction values of $0.5$
is of the order of $10 \%$.
The magnitude of it is compatible with the corresponding deviation in the quasi-particle 
tunneling experiments in the $QHE$ quoted by Gattli \cite{Glattli}, ($2.7-2.65$ vs $3$ ).

We therefore conclude that the low-lying $EFT$ for 
the Mott materials in the tunneling experiments considered
here yields a consistent $(2+1)$-dimensional structure composed of two Chern-Simons fields.
For each Chern-Simons theory, the excitations are described by two classes of Wilson loops 
(vortices): charged and neutral 
modes, of opposite circulation. We speculate that in the low temperature phase, these 
vortices array themselves into
an ordered ($AF$) lattice \cite{Botte-Z-Mott1}. 

It is worthwhile to mention that the $EFT$ \cite{Botte-Z-Mott1} is only the simplest that may 
be proposed to
describe the Mott-class materials. Actually, it simply generalizes to a topological $(2+1)$-dimensional 
theory the underlying $SU(2)$ symmetry of the original fermionic model. The $EFT$ 
contains the minimal number of degrees of freedom at large distance scales
consistent with the original symmetry of the small distance degrees of freedom.
By considering microscopic lattice fermion models with additional degrees of freedom as a starting point, 
it is natural to expected that 
further $EFT$s with enlarged symmetries and field content may be obtained.

We would like to conclude with some remarks. The assumption of planarity is not fundamental, since
the $EFT$ is a Chern-Simons topological field theory that may be defined in $3D$ Euclidean space and quantized 
choosing any time surface \cite{Witten} \cite{zemba}. 
The $EFT$ is the precise field theory describing 
the universality class of the fermionic model (\ref{Model-Ferm-2d})
at the critical point, including in particular all the states in the Hilbert space and  may be, therefore, regarded as a good starting point for further studies along the lines discussed here.
We stress that the relation between the $EFT$ and specific models or characteristics relevant to specific
materials pertains only their universal, long-distances emerging universal data \cite{Polchinski}.
We also remark that the prediction of the universal tunneling exponent $m$ was obtained by a scaling
argument, which is a first-order result in Renormalization Group perturbation theory. 
By the same reason, the zero-width form-factor for the tunneling barrier assumed in (\ref{tunnelings})
does not modify the first-order prediction of $m$, since finite barrier widths would modify (\ref{tunnelings})
by higher order corrections in the transferred momentum, and the $EFT$ considers the limit of vanishing momenta.

\def\RMP{{\it Rev. Mod. Phys.\ }}
\def\PRL{{\it Phys. Rev. Lett.\ }}
\def\CMP{{it Commun.Math.Phys.\ }}
\def\PL{{\it Phys. Lett.}}
\def\PR{{\it Phys. Rev.  \ }}
\def\NP{{\it Nuclear. Phys.}}
\def\PRB{{\it Phys. Rev. B  \ }}
\def\IJMP{{\it Int. J. Mod. Phys.}}

\end{document}